\documentclass{ws-procs975x65}
\usepackage{subfigure}

\def\be{\begin{equation}}
\def\ee{\end{equation}}
\def\bea{\begin{eqnarray}}
\def\eea{\end{eqnarray}}

\def\R{{{\cal{R}}}}
\def\P{{{\cal{P}}}}

\def\vp{{\varphi}}

\def\H{{\cal H}}
\def\cs2{c_{\rm{s}}^2}

\def\U0{{\bar U_0}}

\def\N{{\cal{N}}}

\def\bi{\begin{itemize}}
\def\ei{\end{itemize}}

\begin{document}

\title{\uppercase{Quantifying the behaviour of curvature perturbations near Horizon Crossing}}

\author{\uppercase{Ellie Nalson}$^*$ , \uppercase{Ian Huston} and \uppercase{Karim A.~Malik}}
\address{Astronomy Unit, Queen Mary University of London, Mile End Road, London, UK \\ $^*$ Email: e.nalson@qmul.ac.uk}
\author{\uppercase{Adam J.~Christopherson}}
\address{School of Physics and Astronomy, University of Nottingham, University Park, Nottingham, NG7 2RD, UK}

\begin{abstract}
How much does the curvature perturbation change after it leaves the
horizon, and when should one evaluate the power spectrum? To answer
these questions we study single field inflation models numerically,
and compare the evolution of different curvature perturbations from
horizon crossing to the end of inflation. 
We find that e.g.~in chaotic inflation, the amplitude of the comoving
and the curvature perturbation on uniform density hypersurfaces differ
by up to 180 \% at horizon crossing assuming the same amplitude at the
end of inflation, and that it takes approximately 3 efolds for the
curvature perturbation to be within 1 \% of its value at the end of inflation.
\end{abstract}


\bodymatter

\section{Introduction}
\label{sec:introduction}

It is well known that the curvature perturbations on both uniform density
hypersurfaces, $\zeta$, and on comoving hypersurfaces, $\R$, are conserved on large scales.
The standard approach used to calculate the power spectrum of perturbations
after horizon crossing assumes that the limit $k\to 0$ has been reached, see Ref. \cite{Copeland:1993jj}.
However, immediately after horizon crossing the wavenumber will not yet have become sufficiently small
for this limit to be accurate and gradient terms will still play a role.
Exactly how long this evolution will last and the size of the errors if the curvature perturbation is evaluated too early
are issues addressed in this work \cite{Nalson:2011gc}.

\section{Equations and Numerics}
\label{sec:equations}

We consider single field inflation in a homogeneous, isotropic, 
Friedmann-Robertson-Walker
background, with scalar perturbations of the field and metric to first
order, and work in flat gauge and Fourier space.  The resulting first
order Klein-Gordon equations are solved numerically by following
Salopek et al. \cite{Salopek:1988qh} Initial conditions for the
background are selected for each potential following Huston
\cite{Huston:2010by} and at early times we assume the Bunch-Davies
vacuum \cite{Salopek:1988qh,Huston:2010by}.  We map the initial scalar
field fluctuations, onto conserved quantities, that remain constant in
the limit $k \to 0$ for adiabatic perturbations. We focus on the
curvature perturbation on uniform density hypersurfaces $\zeta\equiv
\psi+\frac{\H}{\rho_0'}\delta\rho$, and the
comoving curvature perturbation $\R\equiv \psi+\frac{\H}{\vp_0'}
\delta\vp$ (see e.g.~Ref.~\refcite{Malik:2008im}).  These two gauge-invariant curvature
perturbations, are related by a constraint equation and $\zeta + \R$
becomes small on super-horizon scales.  We
show plots for the modes $k_1 = 2.77 \mbox{ x } 10^{-5} Mpc^{-1}$, $k_2 =
2.00 \mbox{ x } 10^{-3} Mpc^{-1}$ (WMAP pivot scale) and $k_3 =
1.45 \mbox{ x } 10^{-1} Mpc^{-1}$.  The results below are for the
potential $U = \frac{1}{2} m^2 \vp ^2$ with $m = 6.32$ x
$10^{-6}M_{\rm{PL}} $ but we also obtain results for $U = U_0 +
\frac{1}{2}m^2\vp^2$, $U = \frac{1}{4}\lambda \vp^4$ and $U = \sigma
\vp^{2/3}$.  For full details see Nalson et al. \cite{Nalson:2011gc}
Our numerical results have been verified with a second numerical
program, pyflation \cite{Huston:2011vt}.

To deduce an analytic expression for the curvature perturbation,
we follow Ref.~\refcite{Lidsey:1995np} The Eq.~(3.34) given there is only valid in the large scale limit, however, it
must be evaluated exactly when the corresponding mode crosses the horizon.

\section{Results}
\label{sec:results}

\begin{figure}[ht!]
  \begin{center}
    \mbox{
      \subfigure[\label{fig:pcR}
Percentage difference between $\P(k)$ evaluated, numerically, at the end of inflation and at each time step, 
plotted for $k_1$ (black, left), $k_2$ (blue, middle)
 and $k_3$ (red, right).]{\scalebox{0.5}{\includegraphics[width=120mm]{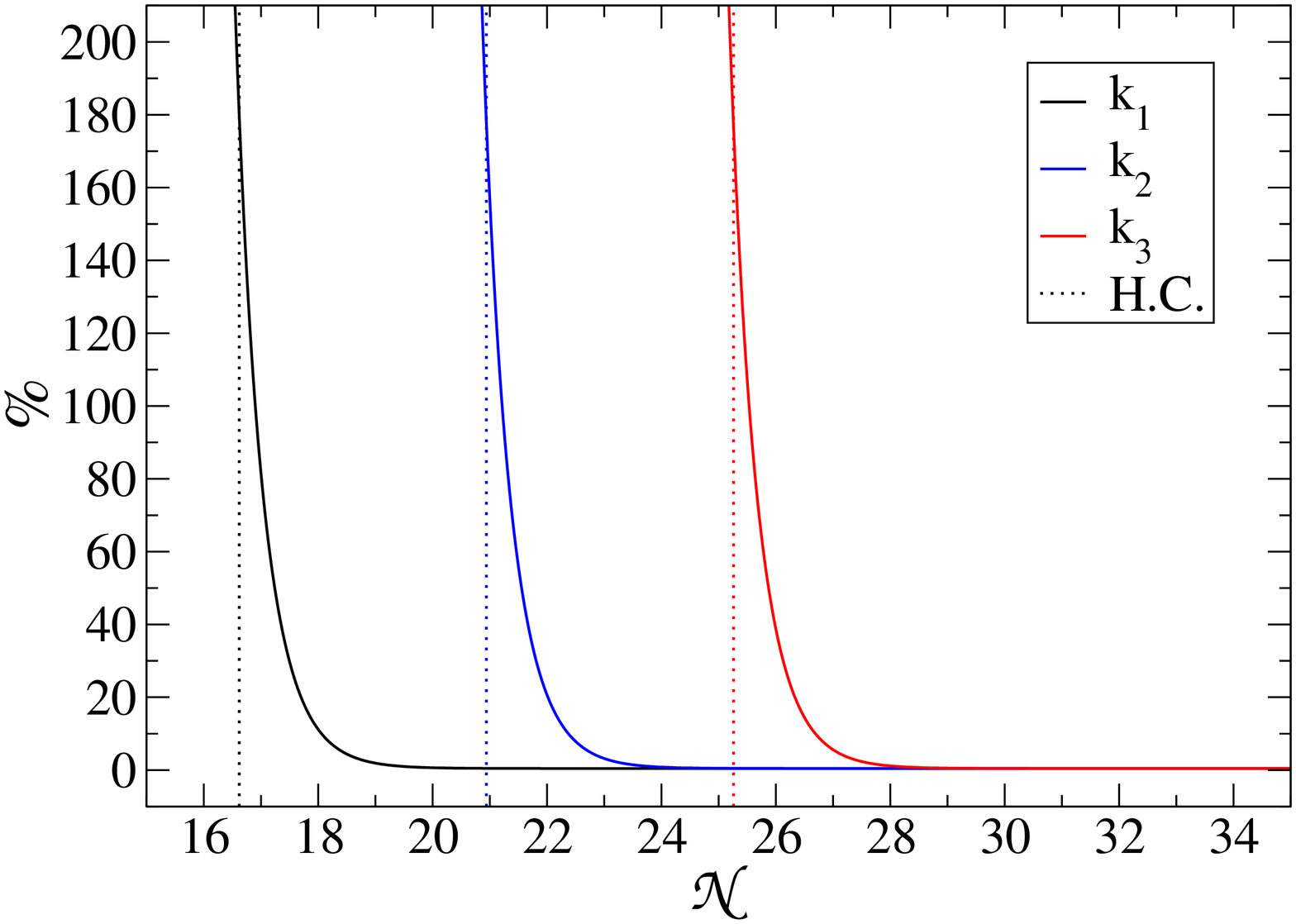}}} \quad
      \subfigure[\label{comparepic2}
 Numerical solutions compared to correct analytic solution and 
na\"ive evaluation of the power spectra at horizon crossing,
plotted for $k_2$ (left lines) and $k_3$ (right lines).] 
{\scalebox{0.5}{\includegraphics[width=120mm]{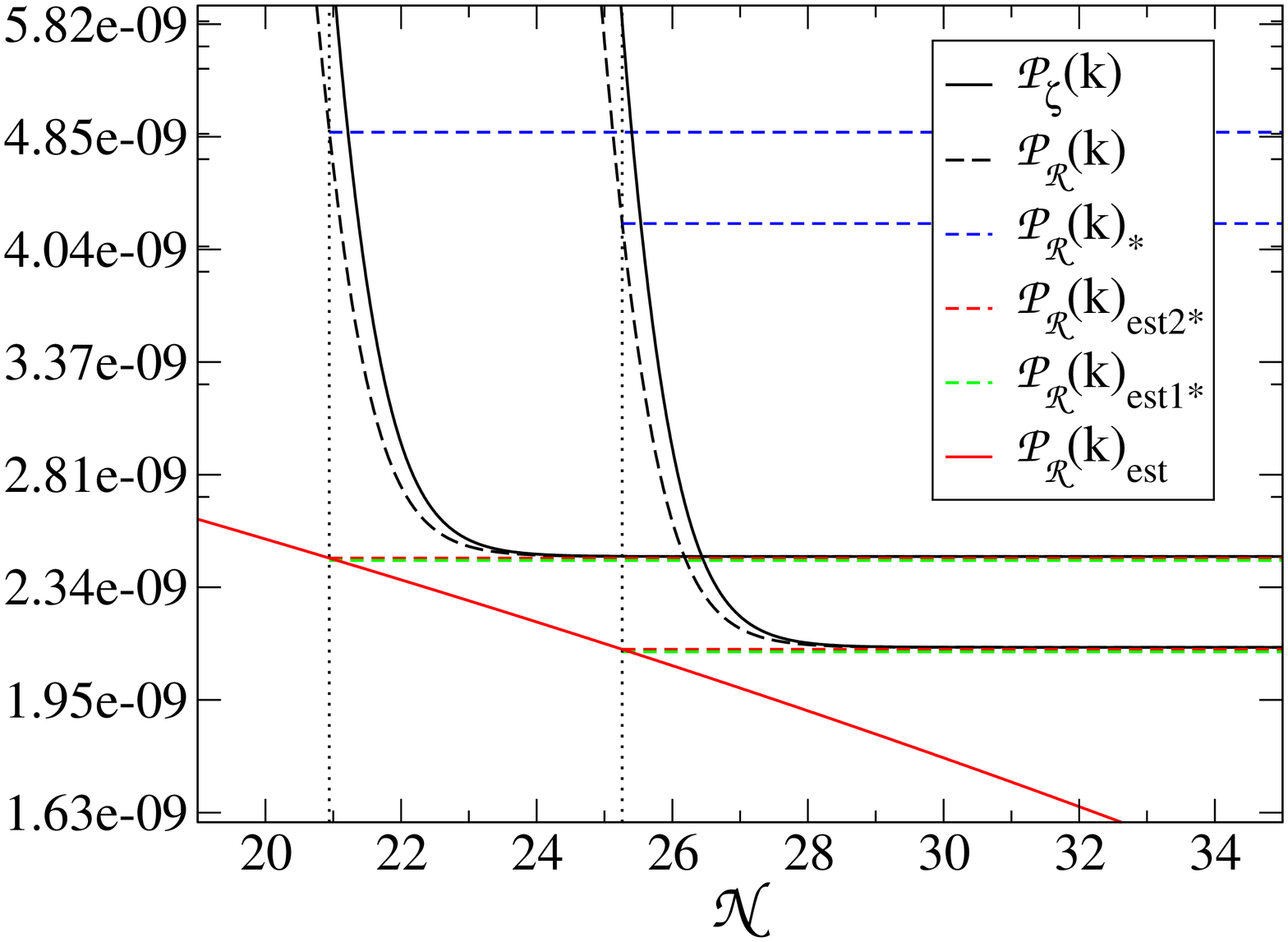}}}
      }
    \caption{The evolution of $\P_{\R}(k)$ and $\P_{\zeta}(k)$ are plotted against the number of efolds, $\N$.}
    \label{fig:compare}
  \end{center}
\end{figure}

As expected, we can see from Fig.~\ref{fig:pcR} that a short time after horizon crossing
there is no longer any appreciable evolution in either the power
spectrum of $\zeta$, $\P_{\zeta}(k)$ or the power spectrum of $\R$,
$\P_{\R}(k)$, but that there is
some evolution immediately after horizon crossing. 
We find that despite $\zeta$ and $\R$ being
equivalent very far outside the horizon, the difference between $|\R|$
and $|\zeta|$ at horizon crossing can be as much as 20\% and $\zeta$ remains significantly larger
for at least a couple of efolds. We also
find that the error in evaluating the power spectra numerically at horizon crossing
rather than using the correct analytic expression or the full
numerical solution at late times can be as much as 180\% for $\P_{\zeta}(k)$ and
100\% for $\P_{\R}$ (this is expected analytically \cite{Polarski:1995jg}).
Lastly we showed that to evaluate
the power spectra without using the analytic expression, one 
would need to wait 3.2 and 2.9 efolds to ensure the
answer for $\P_{\R}(k)$ and $P_{\zeta}$, respectively, are correct to within 1\% of the value at the
end of inflation. Similar results were obtained for the other potentials.

\vspace{-0.5cm}

\vspace{-0.5cm}

In Fig. \ref{fig:compare} we compare the correct
analytic solution and the na\"ive calculation of the power spectrum at
horizon crossing with the numerical solutions.
When we compare the numerical solution to
$\P_{\R}(k)$ evaluated at horizon crossing we find that there is a
100\% error in our answer.
We also see that if one were to use the analytic
expression, but to evaluate it `some efolds
after horizon crossing' rather than at horizon crossing 
one would underestimate the amplitude of the
power spectrum. This corresponds to following the red line in Fig.
\ref{comparepic2} e.g. evaluating the analytic expression 4 efolds after
horizon crossing incurs a 10\% error. 
Lastly we see that the analytic and numerical expressions
do not agree with each other shortly after horizon crossing,
one must wait at least 3.2 efolds for these two values to agree.

\section{Conclusions}
\label{sec:conclusions}

We have quantified the evolution of the curvature
perturbations after inflation and highlighted possible errors which
can occur. As we are entering an era where we hope to constrain
parameters to within a percent using e.g.~Planck data, it is of particular
importance that these errors are both minimised and quantified.
We show the difference between analytic 
and numerical expressions close to the horizon. 
The numerical results, instantaneous values of the power spectrum 
at horizon crossing, while not of observational significance, are useful in many ways, 
~e.g. as initial conditions for other analytical or numerical schemes 
operating outside the horizon. 
If we are interested in late time values we should not evaluate numerical results 
at horizon crossing, as unlike the analytic results they will not be accurate. 
If we are interested in the instantaneous values at or close to 
horizon crossing, the analytic expressions are no longer valid and one must use numerical methods. 
For example, this is important if there is a second phase of evolution which   
starts to dominate during the first three efolds after horizon crossing. 
In conclusion, we have highlighted that confusion between the different curvature 
perturbations, how they are evaluated and when each expression is valid can
introduce additional errors when comparing theoretical results with
observations.

\section*{Acknowledgements}

EN is funded by a STFC studentship. AJC is funded by the Sir Norman
Lockyer Fellowship of the Royal Astronomical Society, IH is funded by
the STFC under Grant ST/G002150/1, and KAM is supported in part by the
STFC under Grants ST/G002150/1 and ST/H002855/1.


\bibliographystyle{ws-procs975x65}
\bibliography{Curvepert}

\end{document}